\documentstyle[12pt]{article}

\title{
\begin{flushright}
{\normalsize Yaroslavl State University\\
             Preprint YARU-HE-93/03\\
             hep-ph/9404289} \\[1cm]
\end{flushright}
The radiative decay of a high energy neutrino in the Coulomb
field of a nucleus}

\author{A.A.Gvozdev, N.V.Mikheev and L.A.Vassilevskaya\thanks{E-mail:
physteo@univ.yars.free.net}\\
{\small\it Division of Theoretical Physics, Department of Physics,}\\
{\small\it Yaroslavl State University, Sovietskaya 14, Yaroslavl 150000,}\\
{\small\it Russia}}

\date{6 December 1993}

\begin{document}

\maketitle

\begin{abstract}

In the framework of the Standard Model with lepton mixing the radiative
decay $\nu_i \rightarrow \nu_j \gamma$ of a neutrino of high ($E_\nu \sim
100 \, GeV$) and super-high ($E_\nu \ge 1 \, TeV$) energy is investigated
in the Coulomb field of a nucleus. Estimates of the decay probability and
``decay cross-section'' for neutrinos of these energies in the electric
field of a nucleus permit one to discuss the general possibility of carrying
out a neutrino experiment, subject to the condition of availability in the
future of a beam of neutrinos of these high energies. Such an experiment
could give unique information on mixing angles in the lepton sector of
the Standard Model which would be independent of the specific neutrino
masses if only the threshold factor ($1 - m_j^4 / m_i^4$) was not close
to zero.

\end{abstract}

\newpage

At the present time the physics of the massive neutrino is the subject of
intensive experimental and theoretical investigations. Given the neutrino's
non-degenerate mass spectrum, it is natural to expect lepton mixing, similar
to the well-known quark mixing, in the lepton sector of the electroweak
theory. Neutrino oscillations [1] are the main source of information on the
mixing angles rigidly correlate with the neutrino's mass spectrum
(see, for example, [2]).
In this work we come up with yet another source of information on the lepton
mixing angles -- namely, the radiative decay of a high-energy neutrino in
an external electromagnetic field which has virtually no correlation with the
mass spectrum. Such an intensive field may be represented, for example,
by Coulomb field of a nucleus.

In our recent paper [3] we studied the neutrino radiative decay $\nu_i
\rightarrow \nu_j \gamma$ in an external magnetic field within the Standard
Model with lepton mixing. The effect of significant enhancement of the
probability of the neutrino radiative decay by a magnetic field (magnetic
catalysis) was discovered. It is important that the probability of the
ultrarelativistic neutrino radiative decay in  external field was found
to be practically independent of the mass of the decaying neutrino, if only
the decay channel is open ($m_i^2 \gg m_j^2$). The result we obtained for
the amplitude of the ultrarelativistic neutrino radiative decay in an uniform
magnetic field can be applied to handle the high-energy neutrino radiative
decay in the Coulomb field of nucleus. Indeed, in this case the Coulomb
field in the decaying neutrino rest frame appears, as does the magnetic field,
very
close to the crossed field ($\vec{\cal E} \perp \vec B$, ${\cal E} = B$).
Here the non-uniformity of the electric field $\vec{\cal E}$ of the nucleus
is of no consequence, because the loop process involved is ``local'' with
the characteristic loop dimension $\Delta x \le (E_\nu e {\cal E})^{-1/3}$
being significantly less than the nucleus size at neutrino energy $E_\nu
\ge 100 \, GeV$. In this case the decay amplitude we have obtained earlier
(see Eq.(8) in Ref.[3]) can be brought into a more convenient form:

\begin{eqnarray}
{\cal M} & \simeq & \frac{e^2 G_F}{\pi^2} \; (\varepsilon^* \tilde F p)
\left[ (1-x) + \frac{m_j^2}{m_i^2} (1+x) \right]^{1/2} \sum_\ell
K_{i \ell} K_{j \ell}^* \, J(\chi_\ell) , \nonumber \\
J(\chi_\ell) & = & i \int\limits_0^1 dt \; z_\ell \int\limits_0^\infty du \;
\exp \left[ - i \, (z_\ell u + u^3 / 3) \right] , \\
z_\ell & = & 4 \left[ (1+x) (1-t^2) \left( 1 - \frac{m_j^2}{m_i^2} \right)
\chi_\ell \right]^{-2/3} , \nonumber
\end{eqnarray}

\noindent where $F_{\mu \nu}$, $\tilde F_{\mu \nu} = \frac{1}{2}
\varepsilon_{\mu \nu \alpha \beta} F_{\alpha \beta}$ are the external
electromagnetic field (Coulomb in our case) tensor and its dual tensor,
$p_\mu$, $m_i$ are the 4-momentum and the mass of the initial neutrino, $m_j$
is the mass of the final neutrino, $\varepsilon_\mu$~is the polarization
4-vector of the photon, $\chi_\ell = \sqrt{e^2 (pFFp)} / m_\ell^3$ is the so
called dynamic parameter, $m_\ell$ is the mass of the virtual charged lepton,
$K_{i \ell}$~is the lepton mixing unitary matrix ($\ell = e, \mu, \tau$),
$x = \cos \theta$, $\theta$~is the angle between the vector $\vec p$
(momentum of the decaying ultrarelativistic neutrino) and $\vec q_0$
(momentum of the photon in the decaying neutrino rest frame). When analyzing
the amplitude (1), one should keep in mind that in the ultrarelativistic
limit the kinematics of the decay $\nu_i(p) \rightarrow \nu_j(p') +
\gamma(q)$ involves nearly parallel 4-momenta $p$, $p'$ and $q$. Thus,
the 4-vector of the neutrino current $j_\alpha = \bar \nu_j(p') \gamma_\alpha
(1+\gamma_5) \nu_i(p)$ is also proportional to those vectors ($j_\alpha
\sim p_\alpha \sim q_\alpha \sim {p'}_\alpha$). It is worth noting that
the amplitude (1) is a sum of three loop contributions ($\ell = e, \mu,
\tau$), each one being characterized by its ``field form-factor''
$J(\chi_\ell)$, which depends on the dynamical parameter $\chi_\ell$.
In an electric field this parameter can be represented as follows:

\begin{equation}
\chi_\ell \simeq \left( \frac{E_\nu}{m_\ell} \right) \,
\left( \frac{e {\cal E}}{m_\ell^2} \right) \, \sin \alpha ,
\end{equation}

\noindent where $\alpha$ is the angle between the vector of the momentum
$\vec p$ of the decaying neutrino and the electric field strength
$\vec{\cal E}$.
In a general way, cumbersome numerical calculations are required to find
the probability. In the limit of super-high neutrino energies ($E_\nu
\ge 1 \, TeV$), however, the situation is drastically simplified, as at
such neutrino energies the conditions $\chi_e \gg \chi_\mu \gg 1$,
$\chi_\tau \ll 1$ are fulfilled in the vicinity of the nucleus.
Therefore, it is sufficient for us to know only the asymptotics of the
function $J(\chi)$:

\begin{eqnarray}
J(\chi) & = & 1 + O(\chi^2), \qquad \chi \ll 1, \nonumber \\
J(\chi) & = & O(\chi^{-2/3}), \qquad \chi \gg 1,
\end{eqnarray}

\noindent so that the amplitude (1) is dominated by the contribution due
to the virtual $\tau$-lepton:

\begin{eqnarray}
K_{ie} K_{je}^* \, J(\chi_e) + K_{i \mu} K_{j \mu}^* \, J(\chi_\mu) +
K_{i \tau} K_{j \tau}^* \, J(\chi_\tau) \simeq K_{i \tau} K_{j \tau}^* .
\nonumber
\end{eqnarray}

\noindent This permits employing the expression for the decay probability
we have obtained earlier (see Eq.(13) in Ref.[3]) which can be written in
the following form:

\begin{equation}
w \simeq \frac{\alpha}{4 \pi} \; \frac{G_F^2}{\pi^3} \; E_\nu e^2
{\cal E}^2 \sin^2 \alpha \; \left( 1 - \frac{m_j^4}{m_i^4} \right) \;
| K_{i \tau} K_{j \tau}^* |^2 .
\end{equation}

It is intriguing to compare this expression for the probability with the
well-known expression for the probability of the radiative decay $\nu_i
\rightarrow \nu_j \gamma$ of a high-energy neutrino in vacuum
(see, for example, [4]):

\begin{equation}
w_0 \simeq \frac{27 \alpha}{32 \pi} \; \frac{G_F^2}{192 \pi^3} \;
\frac{m_i^6}{E_\nu} \; \left( \frac{m_\tau}{m_W} \right)^4 \;
\left( 1 - \frac{m_j^4}{m_i^4} \right) \;
\left( 1 - \frac{m_j^2}{m_i^2} \right) \;
| K_{i \tau} K_{j \tau}^* |^2 .
\end{equation}

\noindent The comparison of the formula demonstrates the enhancing influence
of the external field on the radiative decay

\begin{equation}
R \equiv \frac{w}{w_0} \simeq \frac{512}{9} \;
\left( \frac{m_W}{m_\tau} \right)^4 \;
\left( \frac{E_\nu}{m_i} \right)^2 \;
\left( \frac{e {\cal E}}{m_i^2} \right)^2 \; \sin^2 \alpha .
\end{equation}

As an illustration, let us give a numerical estimate of $R$ in the case
of the decay of a neutrino of energy $E_\nu \sim 1 \, TeV$ in the vicinity
of a nucleus with the atomic number $Z \sim 20$:

\begin{equation}
R \simeq 2 \cdot 10^{+61} \, \left( \frac{1 \, eV}{m_i} \right)^6 \;
\left( \frac{E_\nu}{1 \, TeV} \right)^2 .
\end{equation}

The neutrino radiative decay in substance must result in $\gamma$-quantum
of energy $E_\gamma \sim E_\nu$ being observed as the decay product.
In experiment, this process would appear as inelastic scattering of the
neutrino on the nucleus. Using the expression (4) for the probability and
taking the nucleus as a uniformly charged sphere of radius $r_N$ we obtain
the following expression for the ``cross-section'' of the decay of a
super-high-energy neutrino ($E_\nu \ge 1 \, TeV$) in the electric field of
a nucleus with the atomic number $Z$:

\begin{eqnarray}
\sigma & \simeq & \frac{4}{5} \, Z^2 \, \left( \frac{\alpha}{\pi} \right)^3 \;
\frac{G_F^2 E_\nu}{r_N} \left( 1 - \frac{m_j^4}{m_i^4} \right) \,
| K_{i \tau} K_{j \tau}^* |^2 \nonumber \\
& \simeq & 10^{-44} \, Z^2 \, \left( \frac{10^{-12}cm}{r_N} \right) \,
\left( \frac{E_\nu}{1 \, TeV} \right) \, | K_{i \tau} K_{j \tau}^* |^2
\quad (cm^2).
\end{eqnarray}

\noindent A cumbersome numerical calculation of the probability at more
realistic energies (e.g., at $E_\nu \sim 100 \, GeV$, for which $J(\chi_\mu)$
is not small) shows that Eq.(8) in this case also gives a correct estimate
of the ``cross-section'' within an order of magnitude. This requires the
following substitution in Eq.(8):

\begin{equation}
| K_{i \tau} K_{j \tau}^* |^2 \longrightarrow \sin^2 \Theta_{12} \;
\cos^2 \Theta_{12} \; \overline{| J(\chi_\mu) |^2} ,
\end{equation}

\noindent where $\Theta_{12}$ is the mixing angle of $\nu_1$ and $\nu_2$
(analogous to the Kabibbo angle in the quark sector) and $\overline{|
J(\chi_\mu) |^2} \sim 1$ is the average of the modulus squared of the
``field form-factor'' of the muon loop.

It is worthwhile noting that the ``cross-section'' (8)-(9) we have presented
is, of course, numerically small, but not so small as not to allow discussion
concerning the possibility of such an experiment in the future.
The ``cross-section'' measurement at this level of accuracy would supply unique
information on mixing angles in the lepton sector of the electroweak theory
independent of the specific neutrino masses, if only the threshold factor
$(1 - m_j^4 / m_i^4)$ was not close to zero.

This work was financed in part by a grant of the American Physical Society.

\end{document}